# Data Lakes, Commons and Clouds: A Review of Platforms for Analyzing and Sharing Genomic Data

Robert L. Grossman
University of Chicago

Abstract: Data commons collate data with cloud computing infrastructure and commonly used software services, tools and applications to create biomedical resources for the large-scale management, analysis, harmonization, and sharing of biomedical data.  Over the past few years, data commons have been used to analyze, harmonize and share large scale genomics datasets.   Data ecosystems can be built by interoperating multiple data commons.  It can be quite labor intensive to curate, import and analyze the data in a data commons.  Data lakes provide an alternative to data commons and simply provide access to data, with the data curation and analysis deferred until later and delegated to those that access the data.   We review software platforms for managing, analyzing and sharing genomic data, with an emphasis on data commons, but also covering data ecosystems and data lakes.

**The Challenges of Large Genomic Datasets**

The commoditization of sensors has resulted in new generations of instruments that produce large datasets that are available to genetics researchers. Next generation sequencing produced whole exome and whole genome datasets that were 200 GB to 800+ GB in size, and large projects such as The Cancer Genome Atlas (TCGA) [1] contain more than 2 PB of data and derived data.

Over the next few years, the research community will collect single cell atlases [2], next generation imaging that captures the cellular micro environment, and atlases about the cancer cells interaction with the immunological system, all of which will produce ever larger datasets.

The accumulation of all this data has resulted in several challenges for the genetics research community. First, the size of the datasets is too large for all but the largest research organizations to manage and analyze. Second, the current model in which research groups set up their own computing infrastructure, download their own copy of the data, add their own data, and analyze the integrated dataset is simply too expensive for the government and private funding organizations to support. Third, the IT expertise to set up the required large-scale computing environments and the bioinformatics expertise to set up the required bioinformatics environments is difficult for most organizations to support. Fourth, because of **batch effects** (see Glossary) [3], it is usually considered wise to re-analyze all of the data (from raw data) using a common set of bioinformatics pipelines to minimize the presence of batch effects.

The importance of the appropriate data and computing infrastructure to create "knowledge bases" and "knowledge networks" to support precision medicine has been described in several reports [4, 5].

In this review, we describe some of the data, analysis and collaboration platforms that have emerged to deal with these challenges.

**Platforms for Data Sharing**

*Cloud Computing.* Over the past 15 years, large-scale internet companies, such as Google, Amazon and Facebook, have developed new computing infrastructure for their own internal use that became known as cloud computing platforms [6]. Some of these companies then made these platforms available to customers, including Amazon's Amazon Web Services (AWS), Google's Cloud Platform (GCP), and Microsoft's Azure. Importantly, open source versions of some these platforms were also developed [7], including OpenStack (openstack.org) and OpenNebula (opennebula.org), which enabled organizations to set up their own on-premise clouds. On-premise clouds are also called private clouds [8] to distinguish them from commercial public clouds that are used by multiple organizations.

**NIST** (see Glossary) has developed a definition of cloud computing, which includes the following characteristics [8]:

- elastic in the sense that large-scale resources are available;
- self-provisioned in the sense that a user can provision the computing infrastructure required directly through a portal or **API** (see Glossary).

Although it took time for cloud computing to be adapted for biomedical informatics, there was early recognition within the cancer community of the importance of this technology [9], [10], and several university and institute-based projects developed production level cloud computing platforms to support the cancer research community, including the Bionimbus Protected Data Cloud [11], the Galaxy Cloud [12, 13], Globus Genomics [14], and the Cancer Genome Collaboratory [15]. In addition, commercial companies, including DNAnexus [16] and Seven Bridges [17], also developed cloud-based solutions for processing genomic data.

It may be helpful to divide computing platforms supporting biomedical research into three generations: 1) databases, 2) data clouds, and 3) data commons. See Figure 1.

*Databases and Data Portals.* First generation platforms operated databases in which biomedical datasets were deposited, beginning with GenBank [18]. As the web became the dominant infrastructure for collaboration, **data portals** (see Glossary) emerged as applications that made the data in the underlying databases readily available to researchers. For the purposes here, one can think of a *data portal* as a website that provides interactive access to data in an underlying database. Although data portals are outside the scope of this review, it is still important to mention the UCSC Genome Browser [19] and cBioPortal [20] as some of the most important examples from this category.

The UCSC Genome Browser has been in continuous development since it was first launched in 2000 to help visualize the first working draft of the human genome assembly [19]. Today it contains over 160 assemblies from over 90 species and can be run not only over the web but downloaded and run locally using a version called genome browser in a box (GBiB) [21].

The cBioPortal for Cancer Genomics [20] is a widely used resource that integrates and visualizes cancer genomic data, including mutations, copy number variation, gene expression data, and clinical information. Currently, cBioPortal includes data from TCGA that is processed by Broad's Firehose and data from the International Cancer Genomc Consortium (ICGC) that is processed by the PANCAN Analysis Working group, plus additional smaller datasets [22]. cBioPortal was one of the first cancer data portals to organize data by genomic alterations, such as mutations, deletions, copy number variation and expression levels, in a way that seemed natural to research oncologists and to tie the alterations back to the original cases in order to support further investigation when desired.

With next generation sequencing, the size of genomics datasets began to grow and large scale computing infrastructure is required to process, manage and distribute data. Several systems were developed to process datasets such as the TCGA. CGHub was developed to host the **BAM** (see Glossary) files from the TCGA project [23] by the University of California Santa Cruz. The Firehose system, developed by the Broad Institute, integrates data from TCGA and processes it using applications from **GATK** (see Glossary) [24] and applies algorithms such as **GISTIC2.0** (see Glossary) [25] and **MutSig** (see Glossary) [26]. The results can be browsed and accessed via a website (gdac.broadinstitute.org) and are available in Broad's FireCloud system [27].

*Data Clouds.* Second generation systems colocate computing with biomedical data enabling researchers to compute over the data. A good example of this is the BLAST service [28] provided by NCBI. Over the past decade cloud computing has enabled the colocation of on-demand large scale computing infrastructure that has created new opportunities for the large-scale analysis of hosted biomedical data. In this paper, we use the term **data cloud** (see Glossary) for this integrated infrastructure. A working definition of a *data cloud* for biomedical data is a cloud computing platform [6] that manages and analyzes biomedical data and, usually, integrates the security and compliance required to work with controlled access biomedical data, such as germline genomic data. Examples of biomedical data clouds includes the Bionimbus Protected Data Cloud developed by the University of Chicago [11], the Cancer Genomics Cloud developed by Seven Bridges Genomics [29], the Cancer Collaboratory developed by the Ontario Institute for Cancer Research [30], and the Galaxy Cloud [12, 13] developed by the Galaxy Project.

We now describe three important milestones in the use of large scale cloud computing in genomics. The first was the launch of the NCI Genomics Data Commons [31] that used an OpenStack based private cloud to analyze and harmonize genomic and associated clinical data from over 18,000 cancer tumor-normal pairs, including TCGA [1]. By **data harmonization** (see Glossary) we mean applying a uniform set of pipelines for cleaning, applying quality control criteria, processing, and post-processing submitted data [32]. The second was the development of the three NCI Cloud Pilots: the ISB Cancer Genomics Cloud by the Institute for Systems Biology [33], FireCloud by the Broad Institute [27], and the Cancer Genomics Cloud by Seven Bridges Genomics [29], each of provided cloud based computing infrastructure to analyze TCGA data. The first two used GCP and the third used AWS. A third important milestone was the analysis of 280 whole genomes using multiple distributed public and private clouds by the PANCAN Analysis Working Group [30].

Cloud computing is widely used today to support scientific research for many disciplines outside of the biomedical sciences. In general, the architecture for these systems is simpler since the security and compliance infrastructure required for working with controlled access biomedical data is not required.

*Data Commons.* Third generation systems integrate biomedical data, computing and storage infrastructure, and software services required for working with data to create a **data commons** (see Glossary). Some examples of data commons and six core requirements for data commons are reviewed in [34]. A working definition of a *data commons* is *the colocation of data with cloud computing infrastructure and commonly used software services, tools & applications for managing, integrating, analyzing and sharing data that are exposed through APIs to create an interoperable resource* [34].

Some of the core services required for a data commons include:

*Data common services:*

1. Authentication services for identifying researchers;
2. Authorization services for determining which datasets researchers can access;
3. Digital ID services for assigning permanent identifiers to datasets and accessing data using these IDs ;
4. Metadata services for assigning metadata to a digital object identified by a Digital ID and accessing the metadata;
5. Security and compliance services so that data commons can support controlled access data;
6. Data model services for integrating data with respect to one or more data models.
7. Workflow services for executing bioinformatics pipelines so that data can be analyzed and harmonized.

Accessing controlled access data requires Services 1 and 2. With Service 3, data stored in commons is findable and accessible. With Service 4, data stored in data commons can be reusable and interoperable. In practice, for data to be reusable depends in large part on the quality of the data annotation prepared by the data submitter. With Services 3 and 4, data stored in commons is findable, accessible, reusable and interoperable, which is sometimes abbreviated as FAIR. The importance of making biomedical data FAIR has been stressed in efforts such as the European FORCE11 Initiative [35] and the US NIH BD2K Initiative [36]. Recently, a framework for metrics to measure the "FAIRness" of services has also been developed [37].

Workflow Services 7 in data commons are quite varied and include running existing workflows that have been integrated into the commons and can be used to analyze data in the commons, pulling existing workflows from workflow repositories outside the commons and applying them to data in the commons, and developing new workflows and using them to analyze data in the commons. Also, some commons allow users to execute workflows, while others limit this to the data commons administrators.

An example of a data commons is the NCI Genomic Data Commons (GDC) [31, 38], was used by over 100,000 distinct cancer researchers in 2017. With the GDC [31], data commons began to *curate and integrate* contributed data using a common data model (core service 6), *harmonize* the contributed data using a common set of bioinformatics pipelines (core service 7), support the visual *exploration* of data through a data portal, and *expose* APIs to the core services 1)-5) to support third party applications over the integrated and harmonized data.

*Project data: object data and core data.* Data in a data commons is usually organized into projects, with different projects potentially having their own data model and collecting different subsets of clinical, molecular, imaging and other data. It is an open question of how best to organize data across projects so that it can integrated, harmonized and queried. One natural division that is emerging is the distinction between the **structured data** (see Glossary), the unstructured data, and the **data objects** (see Glossary) in a project. The *object data* typically include FASTQ or BAM files [39] used in genomics, image files, video files, and other large files, such as archive or backup files associated with a project. The *structured data* includes clinical data, demographic data, biospecimen data, variant data [40], and other data associated with a data schema. The unstructured data includes text, notes, articles, and other data that is not associated with a schema.

Part of the curation process is to align the structured data in a project with an appropriate ontology. Examples include using the human phenotype ontology (HPO) [41] and the NCI Thesaurus [42] for curating clinical phenotype data, and CDISC [43] for curating clinical trials data. It can be quite challenging and labor intensive to match ontologies to clinical data and a number of tools have been developed to make this easier [44, 45].

If we call all the all the structured data, unstructured data, and the associated schemas *project core data,* then it is quite common for the project's object data to be 1,000 times (or more) larger than the project's core data. For example, with the TCGA's projects [1], the data objects were measured in 10s-100s TB, while the project's core data were measured in 10s of GB.

In practice, a project's object data are assigned **GUIDs** (see Glossary) and metadata and stored in clouds using services 3 and 4 and are immutable (although new versions may be added to the project), while the project core data is often updated, as part of the curation and QA process and as new data is added to the project.

A project object data is searched via its metadata (Core Service 4), while project core data can be searched via its data model (Core Service 6). Of course, a project's object data can be processed to produce features, which can then be managed and searched. Examples include developing algorithms for identifying particular types of cells in cell images and searching for these cells or processing BAM files to compute data quality scores and searching for BAM files with particular data quality problems. When data is curated and integrated with a common data model, synthetic cohorts can be created through a query, such as, "find all males over 50 years of old that smoked and have a KRAS mutation [46]."

Another way to think of this is that core services 1) - 5) support the "shallow" indexing and search via metadata, while core services 1) – 6) support "deep" indexing and search via the data model attached to project core data. In either case, when the services are exposed via APIs to third party applications, data becomes portable, and data commons become interoperable, both of which are usually thought of as important requirements [34].

*Data Lakes.* Sometimes the term **data lake** (see Glossary) is used when data is stored simply with digital IDs and metadata (shallow indexing), but without a data model. Data models and schemas are used when the data is written or when the data is analyzed, but not when the data is stored. Additional information about data lakes can be found in [47]. Since it can be very labor intensive to import data with respect to a data model, and since not all the data in a commons is used, this has the advantage that the effort to align the data with a data model is not needed until the data is analyzed. Of course, at the time the data is analyzed and aligned with a data model, the expertise to do this may no longer be easily available.

Through the use of cloud computing, data commons can support large scale data, but this also creates sustainability challenges, due to the cost of large scale storage and compute. One sustainability model that can be attractive to an organization is to provide the data at no cost, but to control the cost of the computing resources by using a *pay for compute model* [34], establishing quotas for compute, giving compute allocations, or distributing "chits" that can be redeemed for compute.

Just as data lakes required less curation than data commons, data catalogs required less curation than data lakes. A *data catalog* is simply a listing of data assets, some basic metadata, and their locations, but without a common mechanism for accessing the data, such as used in a data lake.

*Workflows.* Bioinformatics workflows are often data intensive and complex, consisting of several different programs with the outputs of one program used as the inputs to another. For this reason, specialized workflow management systems have been developed so that workflows can be mapped efficiently to different high performance, parallel and distributed computer systems. Workflow languages have been developed so that domain specialists knowledgeable about the workflows can describe the workflows in a manner that is independent of the specific underlying physical architecture of the system executing the workflows.   Despite many years of effort though, there is still no standard language for expressing workflows in general, and bioinformatics workflows in particular [48, 49].   Within the cancer genomics community, the Common Workflow Language (CWL) [50] is gaining in popularity.  The GA4GH Consortium (ga4gh.org) supports a technical effort to standardize bioinformatics workflows, which includes the workflow execution services (WES) and task execution service (TES).  With the growing use of **container** (see Glossary) based environments for program execution, such as **Docker** (see Glossary), it is becoming more common to encapsulate workflows in containers to make them easier to reuse [51].  Prior to the wide adoption of containers, workflows were previously encapsulated in virtual machines for the same reason.  Examples of services for accessing reproducible workflows include Dockstore [52] and Biocompute Objects [53].

**Data and Commons Governance**

A common definition of IT governance is [54]: 1) Assure that the investments in IT generate business value. 2) Mitigate the risks that are associated with IT.  3) Operate in such a way as to make good long-term decisions with accountability and traceability to those funding IT resources, those developing and support IT resources, and those using IT resources.  This definition can be easily adapted to provide a good definition for data commons governance: 1) Assure that the investments in the data commons generate value to the research community. 2) Manage the balance between the risks associated with participant data and the benefits realized from research involving this data [55]. 3) Operate in such a way so as to make good long-term decisions with accountability and traceability to those sponsors that fund the data commons, the engineers that develop, manage and operate the data commons, and the researchers that use it.

An overview of principles for data commons and a description of eight principles for biomedical data commons can be found in [56]. A survey of how data is made available and controlled in commons is in [57]. A survey of data commons governance models is in [58]. The GA4GH framework for sharing data is described in [55].

The data governance structure for international data commons, like the INRG Data Commons [59] and the ICGC Data Commons [60], can be challenging, and may restrictions on the movement of the underlying controlled access genomic data.

**Building and Operating a Data Commons**

Building a data commons usually consists of the following steps:

1. Put in place data governance agreements that govern the contribution, management and use of the data in the data commons and common governance agreements that govern the development, operations, use, and sustainability of the commons.
2. Develop a data model (or data models) that describe the data in the commons.

3. Set up and configure the data commons itself.

4. Work with the community to submit data to the data commons.

5. Import, clean and curate the submitted data.

6. Process and analyze the data using bioinformatics pipelines to produce harmonized data products.  This is often done with analysis working groups.

7. Open up the commons to external researchers, third party applications, and interoperate with other commons.

See Figure 2.

To support the activities, a data commons usually has the following components:

a) A data exploration portal (or more simply a data portal) for viewing, exploring, visualizing and downloading the data in the commons.

b) A data submission portal for submitting data to the commons.

c) An API supporting third party applications.

d) Systems for the large-scale processing of data in the commons to produce derived data products.

e) Systems to support analysis working groups and other team science constructs used for the collaborative analysis and annotation of data in the commons.  What are being called *workspaces* are one of the mechanisms that are emerging to support this.

**Data Ecosystems Containing Multiple Data Commons**

As the number of data commons grow, there will be an increasing need for data commons to interoperate and for applications to be able to access data and services from multiple data commons. It may be helpful to think of this situation as laying the foundation for a data ecosystem [61].

Sometimes the data commons services 1) – 6) described above are called *framework services*, since they provide the framework for building a data commons, and, in fact, can be used to support multiple data commons that interoperate (Figure 3, Key Figure). As mentioned above, when these services are exposed through an Application Programming Interface (API), either as part of a data commons or as a part of framework services supporting multiple data commons, they can support an ecosystem of third party applications [46].

There is no generally accepted definition of a data ecosystem at this time, but, at the minimum, a *data ecosystem* for biomedical data (as opposed to a data commons) should support:

1. Authentication and Authorization services so that a community of researchers can access an ecosystem of data and applications with a common (research) identity and common authorization that is shared across data commons and applications.
2. A collection of applications that are powered by APIs that are FAIR compliant that are shared across multiple data commons.

3. The ability for multiple data commons to interoperate through framework services and, preferably, through data peering [34] so that access to data across data commons and applications is transparent, frictionless and without egress charges, as long as the access is through a digital ID.
4. Shared data models, or portions of data models, to simplify the ability for third party applications to access data from multiple data commons and applications.  Projects within a larger overall program, or in related programs, may share a data model.  More commonly, different projects may share some common data elements within a core data model, with each project having additional data elements unique for that project.
5. Support for workspaces that may include:
    a. The ability to create synthetic (or virtual) cohorts and export cohorts to workspaces;
    b. The ability to execute bioinformatics workflows within workspaces;
    c. Workspace services for processing, exploring and analyzing data using containers, virtual machines or other mechanisms.
6. Security and compliance services.

Often workspace services 5b) and 5c) use a user-pay model as mentioned above.

An example of a cancer data ecosystem is the NCI Cancer Research Data Commons or NCRDC [62].  The NCRDC spans the GDC [31] and the Cloud Resources [62], so that both AWS and the GCP can be used to both analyze data from the GDC as well as to support integrative data analysis across data uploaded by researchers with data from the GDC and other third party datasets. Data commons for proteomic and imaging data are in the process of being added to the NCRDC.   The NCRDC uses the Framework Services described above so that multiple data commons and other NCRDC resources can share authentication, authorization, ID and metadata services.  In particular, this approach allows applications to be built that span multiple data commons.

**Summary and Future Directions**

We have reviewed some of the more recent data and computing platforms that have been used to analyze large scale data being produce in biology, medicine and health care, with a particular emphasis on data commons. See Figure 4 for an overview of the different platforms.   Data commons provide several important advantages, including:

- Data commons support repeatable, reproducible and open research.
- Some diseases are dependent upon having a critical mass of data to provide the required statistical power for the scientific evidence (e.g. to study combinations of rare mutations in cancer)
- With more data, smaller effects can be studied (e.g. to understand the effect of environmental factors on disease).

- Data commons enable researchers to work with large datasets at much lower cost to the sponsor than if each researcher set up their own local environment.
- Data commons generally provide higher security and greater compliance than most local computing environments.
- Data commons support large scale computation so that the latest bioinformatics pipelines can be run.
- Data commons can interoperate with each other so that over time data sharing can benefit from a "network effect"

Over the next few years, one of the most important changes will be the ability of patients to submit their own data to a data commons and to gain some understanding of their own data in terms of the overall data available in the commons and their broader data ecosystem the commons is part of. The ability of patients to contribute their own data and to have control over how the data is used by the research community [63] is an important aspect of what is sometimes called patient partnered research.

**Acknowledgments**

This project has been funded in in part with Federal funds from the National Cancer Institute, National Institutes of Health, Task Order No. 17X053 and Task Order No. 14X050 under Contract No. HHSN261200800001E. The content of this publication does not necessarily reflect the views or policies of the Department of Health and Human Services, nor does mention of trade names, commercial products, or organizations imply endorsement by the U.S. Government.

**Captions:**

Figure 1. This figure shows some of the important differences between data clouds and data commons.

Figure 2. Data commons support the entire life cycle of data, including defining the data model, importing data, cleaning data, exploring data, analyzing data, and then sharing new research discoveries.

Figure 3. This diagram shows how data commons framework services can support multiple data commons and an ecosystem of workspaces, notebooks and applications.

Figure 4. Data platforms can be categorized along four axes: the data architecture, the extent of the data curation and harmonization, the analysis architecture of a resource, and the analysis architecture of the ecosystem.  The red lines can be viewed as classifying platforms using parallel coordinates and these four dimensions.  The top line are the parallel coordinates associated with the NCI Cancer Research Data Commons, the line below are the parallel coordinates for the NCI Genomic Data Commons, the two lines below are two possible architectures for data lakes, while the bottom line is an architecture for a repository of files.

**Glossary**

**Application Programming Interface (API):** An API is a specification for how two different software programmers communicate with each other and an implementation of the specification in computer code.

**BAM**: The Binary Alignment Map (BAM) is a binary format that is widely used for storing molecular sequence data.

**Batch effects:** Batch effects are differences in samples that are the results of differences in laboratory conditions, materials used to prepare the samples, such as reagents, personnel that prepare the samples, and other differences like these. Batch effects are often an important confounding factor in high throughput sequence data.

**Container:** A container for running software is package of software that includes everything needed to run a software application, including the application's code, as well as the runtime environment, system tools, system libraries, configurations and settings. Containers are designed to be run in different types of computing environments with no changes.

**Data clouds:** A data cloud is a cloud computing platform for managing, analyzing and sharing datasets.

**Data commons**: A data commons co-locates data with cloud computing infrastructure and commonly used software services, tools & applications for managing, integrating, analyzing and sharing data that are exposed through APIs to create an interoperable resource.

**Data lake:** A data lake is a system for storing data as objects, where the objects have an associated GUID and (object) metadata, but there is no data model for interpreting the data within the object.

**Data harmonization**: Data harmonization as the process that brings together data from multiples sources and applies uniform and consistent processes, such as uniform quality control metrics to the accepted data; mapping the data to a common data model; processing the data with common bioinformatics pipelines; and post-posting the data using common quality control metrics.

**Data object:** In cloud computing, a data object consists of data, a key, and associated metadata. The data can be retrieved using key and the metadata associated with a specific data object can be retrieved, but more general queries are not support. Amazon's S3 storage system is a widely used storage system for data objects.

**Data portal:** A data portal is a website that provides interactive access to data in an underlying data management systems, such as a database. Data commons, data lakes can also have data portals.

**Docker:**  Docker is a software program for running containers developed by the company Docker, Inc.   The containers it runs are often called Docker ontainers.

**Genome Analysis Toolkit (GATK):**  GATK is a widely used collection of bioinformatics pipelines and associated best practices for variant discovery and genotyping developed by the Broad Institute.

**GISTIC2.0:** GISTIC is a probabilistic algorithm for detecting somatic copy-number alterations (SCNAs) that are likely to drive cancer growth.

**Globally Unique Identifier (GUID):** A GUID is an essentially unique identifier that is generated by an algorithm so that no central authority is needed, but rather different programs running in different locations can generate GUID with a low probability that they will collide.  A common format for a GUID is the hexadecimal representation of a 128 bit binary number.

**MutSig:** MutSig (for mutation significance) is a probabilistic algorithm and associated software application that analyzes a list of mutations produced from DNA sequencing data to identify genes that were mutated more often than expected by chance, given background mutation processes.

**NIST:** National Institute of Standards and Technology is a US federal agency that advances measurement science and develops standards.  NIST has developed definitions in standards for cloud computing and information security.

**Structured data:** Data is structured if it is organized into records and fields, with each record consisting of one or more data elements (data fields).   In biomedical data, data fields are often restricted to controlled vocabularies to make querying them easier.

**Highlights**

Data commons collate data with cloud computing infrastructure and commonly used software services, tools and applications to create biomedical resources for the large-scale management, analysis, harmonization, and sharing of biomedical data.

Data commons support repeatable, reproducible and open research.

Data lakes provide access to a collection of data objects that can accessed via digital IDs and searched via their metadata.

A simple data ecosystem can be built when a data commons exposes an API that can support a collection of third party applications that can access data from the commons.  More complex data ecosystems arise when multiple data commons can interoperate and support a collection of third party applications over a common set of core services (framework services), such as services for authentication, authorization, digital IDs, and metadata.

Reproducibility is a growing concern in biomedical research.  Maintaining datasets over the long term in data commons that can be: 1) accessed via digital IDs and searched via metadata; 2) processed and re-processed using workflows expressed in workflow languages, such as the Common Workflow Language; 3) with software applications encapsulated in containers and virtual machines makes reproducible more likely.

# Outstanding Questions

- In practice, uploading clinical phenotype data into a data commons so that it is aligned with the data common's data model and can be harmonized with the other the data in the commons is quite labor intensive. An open question is how to develop bioinformatics tools and associated frameworks so that data can be transformed automatically or semi-automatically into the proper format.

- Developing software architectures and associated platforms that can that can query and aggregate data from multiple data commons is an important challenge.

- In general, different commons will have both large and small differences in the practices and standards used for assigning clinical phenotype. Developing applications that can query and aggregate data from multiple data commons even when there are minor (or major) differences between the clinical phenotype data is an important challenge.

- In practice, researchers will be analyzing data using applications that are hosted across multiple commercial public clouds, while those operating data commons will try to reduce their costs by focusing on 1 or 2 public or private clouds. What are the software architectures and operating procedures so that data commons can operate across just one or two public or private clouds but support researchers across multiple clouds?

- Moving data projects between data commons is important so that data commons don't begin to "silo" data. What are appropriate serialization formats so that projects can be efficiently imported and exported between data commons?

## Databases

1982 - present

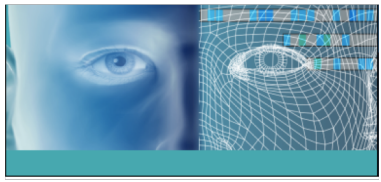

- Data repository
- Researchers download data.

## Data Clouds

2010 - 2020

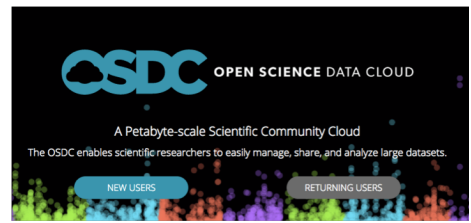

- Supports large datasets & data intensive computing with cloud computing
- Researchers can analyze data with their own virtual workspaces and applications and collaborate with other researchers with collaborative workspaces (data does not have to be downloaded to be analyzed)

## Data Commons

2014 - 2024

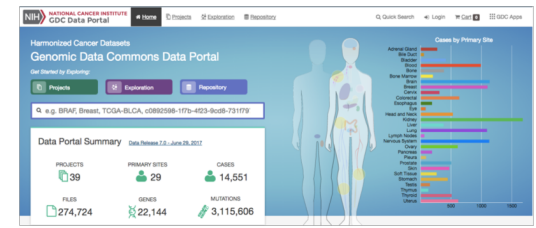

- Supports large datasets and data intensive computing with cloud computing
- Workspaces
- Common data models
- Core data services
- Data & Commons Governance
- Harmonized data
- Data sharing
- Reproducible research

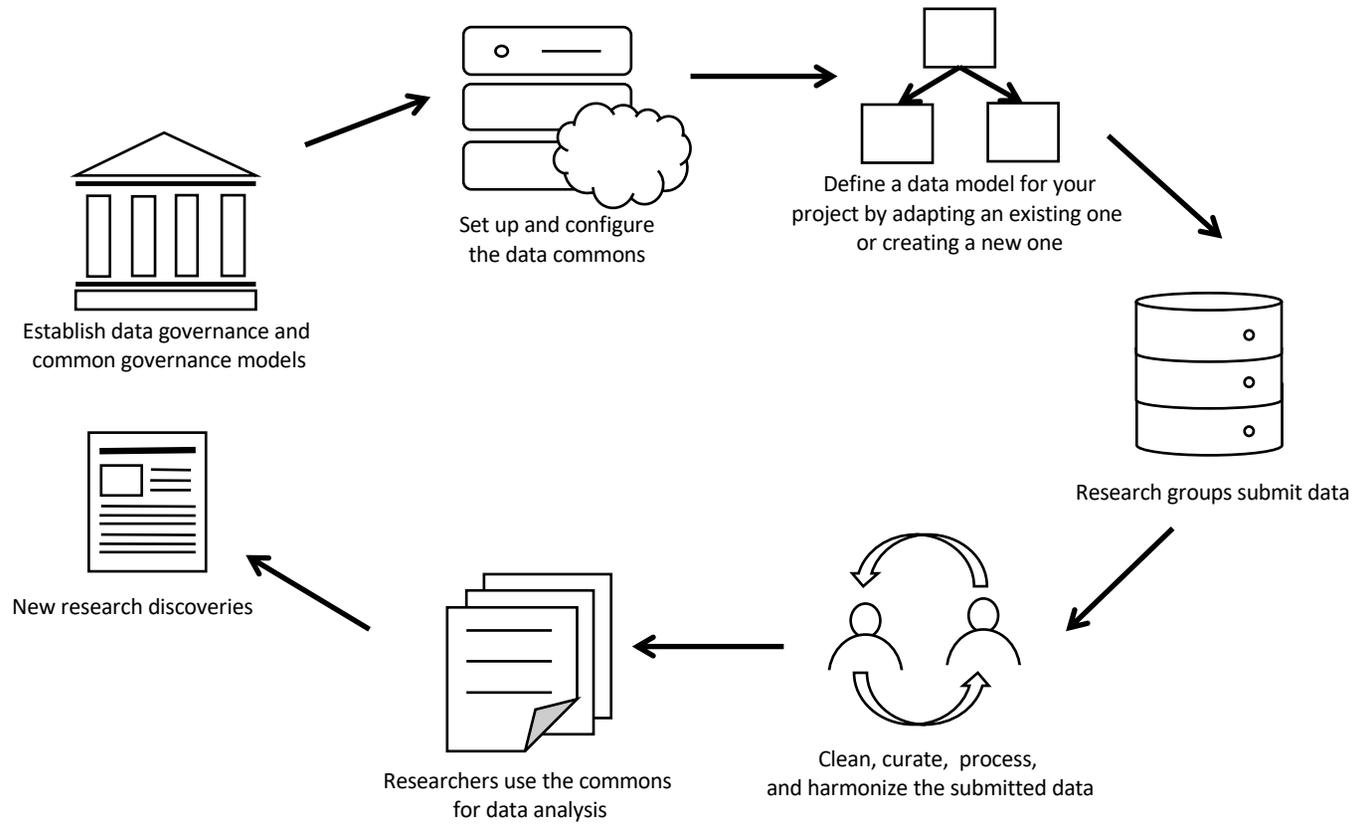

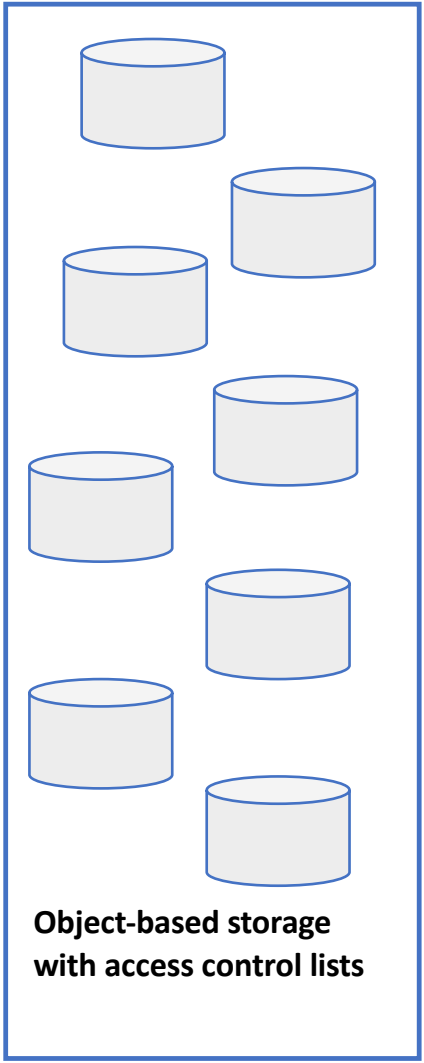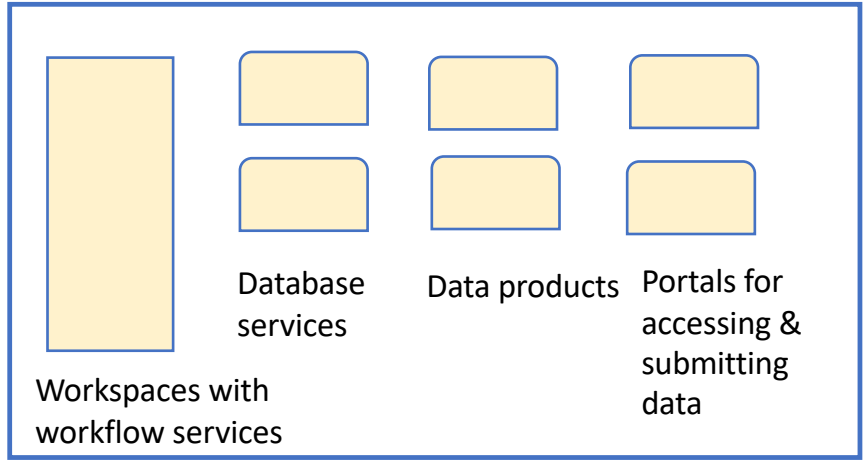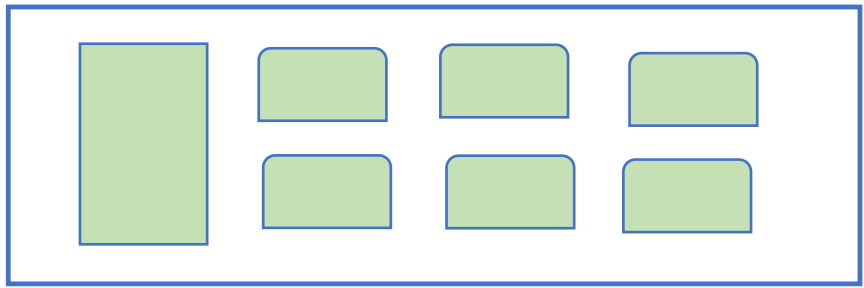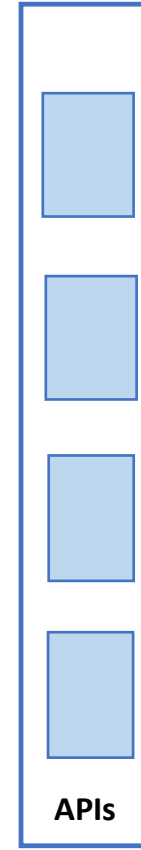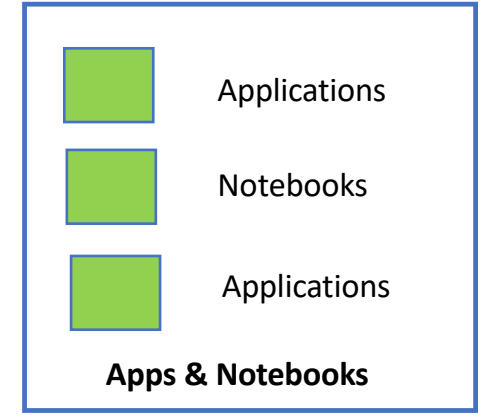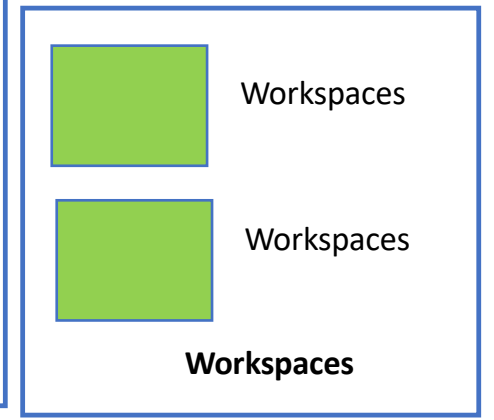

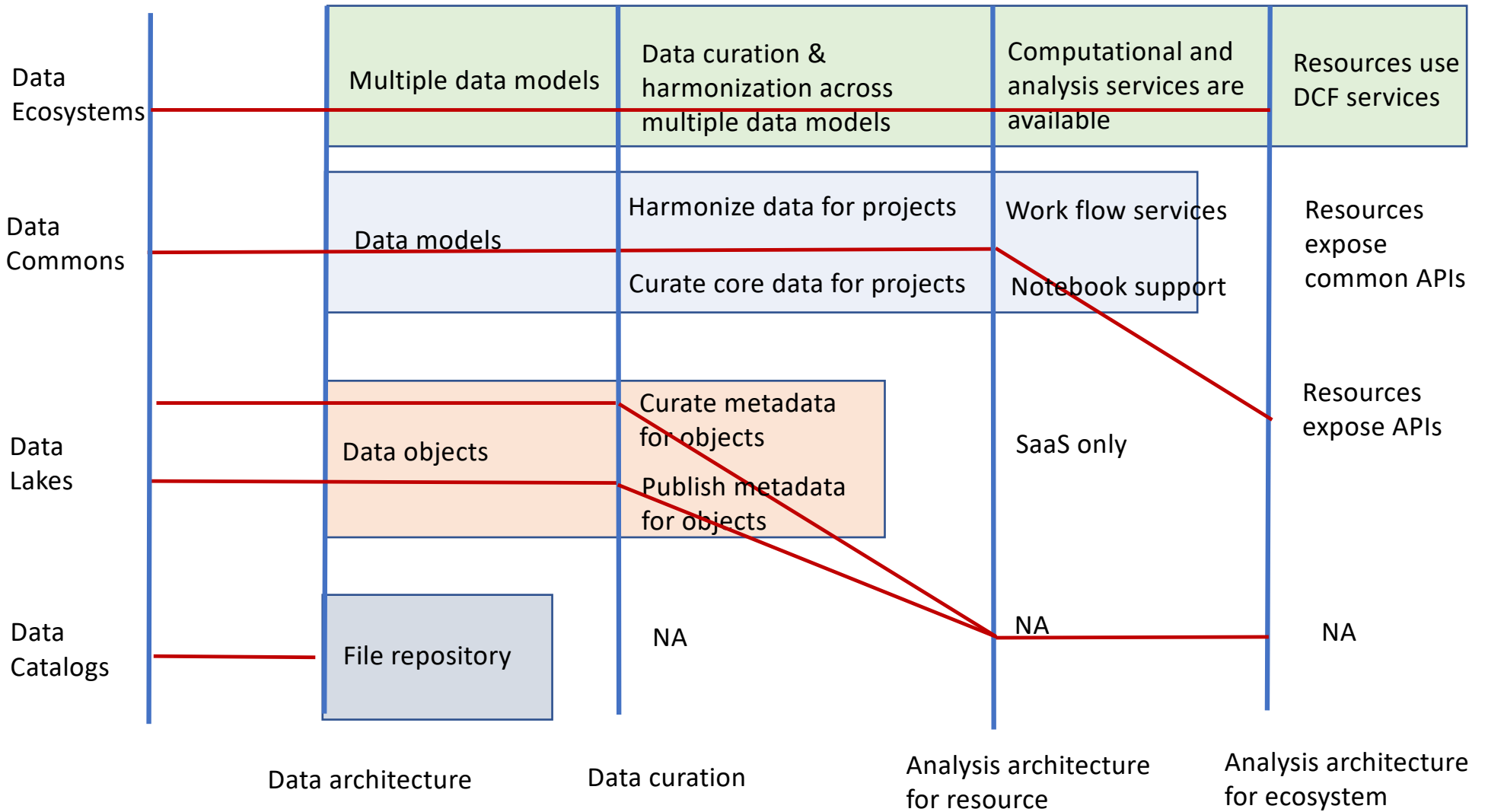